\documentclass[twocolumn,pra,aps,floatfix,superscriptaddress]{revtex4-1}

\usepackage{amssymb,amsmath}
\usepackage{graphicx}
\usepackage{comment}
\usepackage[dvipsnames]{xcolor}
\usepackage[colorlinks,urlcolor=blue,citecolor=blue,linkcolor=blue]{hyperref}
\usepackage{cleveref}
\usepackage{bigints}
\usepackage[export]{adjustbox}

\usepackage{pagecolor,lipsum}

\usepackage[normalem]{ulem}

\newcommand{\q}{\mathbf{q}}

\begin{document}

\title{Long-lived trimers in a quasi-two-dimensional Fermi system}

\author{Emma K.~Laird}
\thanks{Both authors contributed equally to this work.}
\affiliation{School of Physics and Astronomy, Monash University, Victoria 3800, Australia}

\author{Thomas Kirk}
\thanks{Both authors contributed equally to this work.}
\affiliation{School of Physics and Astronomy, Monash University, Victoria 3800, Australia}
\affiliation{London Centre for Nanotechnology, 17-19 Gordon Street, London, WC1H 0AH, United Kingdom}

\author{Meera M.~Parish}
\affiliation{School of Physics and Astronomy, Monash University, Victoria 3800, Australia}

\author{Jesper Levinsen}
\affiliation{School of Physics and Astronomy, Monash University, Victoria 3800, Australia}

\date{\today}

\begin{abstract}
We consider the problem of three distinguishable fermions confined to a quasi-two-dimensional (quasi-2D) geometry, where there is a strong harmonic potential in one direction.  We go beyond previous theoretical work and investigate the three-body bound states (trimers) for the case where the two-body short-range interactions between fermions are unequal.  Using the scattering parameters from experiments on ultracold $^{6}$Li atoms, we calculate the trimer spectrum throughout the crossover from two to three dimensions.  We find that the deepest Efimov trimer in the $^{6}$Li system is unaffected by realistic quasi-2D confinements, while the first excited trimer smoothly evolves from a three-dimensional-like Efimov trimer to an extended 2D-like trimer as the attractive interactions are decreased.  We furthermore compute the excited trimer wave function and quantify the stability of the trimer against decay into a dimer and an atom by determining the probability that three fermions approach each other at short distances.  Our results indicate that the lifetime of the trimer can be enhanced by at least an order of magnitude in the quasi-2D geometry, thus opening the door to realizing long-lived trimers in three-component Fermi gases.
\end{abstract}

\pacs{}

\maketitle

\section{Introduction}
\label{sec:Introduction}

The behavior of three quantum particles interacting with short-range interactions is a fundamental problem in physics that is relevant to a variety of systems ranging from nucleon clusters~\cite{Bedaque2000} to quantum magnets~\cite{nishida2013efimov}.  Our capability to investigate three-body systems has been greatly enhanced by recent advances in the manipulation and cooling of trapped atoms. Here, one can realize a range of cold-atom systems with different quantum statistics and different dimensionalities~\cite{RevModPhys80885}. In all of these scenarios, a key role is played by three-body bound states, i.e., \textit{trimers}, whose existence or otherwise essentially determines the energy spectra and scattering properties.

Of particular interest has been the so-called Efimov effect~\cite{ScatteringConcepts_NaturalLength_Universality_Efimov_Brateen,EfimovEffect_History_Grimm,ShimpeiReview}, which corresponds to a series of trimer states that become infinitely numerous when the short-range interactions are tuned to be resonant. Such an effect was first predicted for three identical bosons more than 40 years ago~\cite{EfimovEffect_Efimov}, and the deepest Efimov trimers have since been observed in atomic Bose gases~\cite{EfimovObservation_Original_Kraemer,huang2014,kunitski2015observation,PhysRevLett.119.143401}, Bose--Fermi mixtures~\cite{PhysRevLett.112.250404,PhysRevLett.113.240402}, and three-component Fermi gases~\cite{FequencyAssociation_EfimovTriemrs_jochim,3-component_BindingEnergyMeasurement_Ueda}.  However, the Efimov trimers observed in the cold-atom system are highly unstable towards decay into deeper bound dimers --- indeed, the trimers are typically detected indirectly via three-body loss resonances at low energy~\cite{EfimovObservation_Original_Kraemer}.  While stable trimers have been produced by using $^4$He atoms~\cite{kunitski2015observation}, this system lacks the tunability of metastable ultracold atomic gases. Thus, it remains an elusive goal to engineer long-lived trimers that can ultimately be used as building blocks for quantum simulators and correlated phases of matter~\cite{rapp2007,3-component_Background_Nishida,kirk2017}.

The short lifetime of Efimov trimers in the cold-atom system is predominantly due to the large weight of the trimer wave function at short distances; i.e., there is a high probability that three particles will approach each other at close range and then decay into a deeply bound molecule and an unbound atom. Therefore, one can enhance the trimer stability by engineering a more spatially extended three-body wave function. Such a scenario can, in principle, be achieved by confining identical bosons to a two-dimensional (2D) plane~\cite{Quasi2D_EfimovTrimers_Parish,3BodyRecombination_2D__Incao}, a geometry which has been realized in Bose-gas experiments using optical lattices~\cite{hadzibabic2006berezinskii,PhysRevLett.102.170401,Hung2011}. In this case, the weakest bound Efimov states are destroyed by the quasi-2D confinement~\cite{Quasi2D_EfimovTrimers_Parish,yamashita2015,Sandoval2017} and, in the 2D limit, one has two spatially extended ``universal'' trimer states that are completely determined by the low-energy 2D scattering parameters~\cite{bruch1979binding}.  However, there are practical difficulties in accessing these 2D-like trimers in the Bose system since one must start from an attractive quasi-2D Bose gas, which is inherently unstable~\cite{huang2018}.

In this paper, we circumvent this problem by considering the trimers formed from a three-component \textit{Fermi} gas.  Such a system can be experimentally realized with $^{6}$Li atoms, since the three lowest hyperfine states of $^{6}$Li all have near-resonant $s$-wave interactions~\cite{3-component_reference_Experimental_FeshbachRsonancesLithium_ScatteringLength_Bartenstein}, and Efimov trimers have already been observed in three dimensions (3D):\,\,\,the ground and first-excited trimer states have been detected indirectly via three-body loss resonances~\cite{3-component_FirstExperiment_Jochim,3-component_RecombinationLoss_Hara,3-component_Efimov_RecombinationLoss2_Hara,Efimov_AtomDimer_Resonance_Jochim,Efimov_AtomDimer_Resonance_Ueda}, while the latter has also been directly accessed via radio-frequency association~\cite{FequencyAssociation_EfimovTriemrs_jochim,3-component_BindingEnergyMeasurement_Ueda}.  On the other hand, it is now standard practice to create stable two-component Fermi gases confined to a 2D plane~\cite{Q2DBackground_Review_Meera}, and this has 
already been achieved with $^6$Li atoms~\cite{From2Dto3D_Experimental_Vale,2Dexperiment_tightconfinement_Lithium6_Zwierlein,3-component_reference_Experimental_Quasi2D_Pairing_Experimental_Jochim,PhysRevLett.117.093601}. Thus, it should be feasible to associate quasi-2D $s$-wave trimers from atoms and dimers in the quasi-2D two-component $^6$Li system, similar to what has been done in 3D~\cite{FequencyAssociation_EfimovTriemrs_jochim,3-component_BindingEnergyMeasurement_Ueda}.  The advantage of this approach is that we only require one species of atoms, in contrast to alternative proposals for stable trimers that require two different types of fermion with a large mass ratio~\cite{KartavtsevMalykh_Trimers_Kartavtsev,LongLiveTrimers_2-component_Jesper,Efimov_KartavtsevMalykh_CrossoverTrimers_Shimpei}.

Using realistic scattering parameters for the three-component $^6$Li system, we compute the spectrum of trimers under different quasi-2D confinements as a function of interaction strength. Since the scattering lengths for the three pairwise interactions are different, we must solve a more involved set of integral equations for the confined system, unlike the case of identical bosons~\cite{Quasi2D_EfimovTrimers_Parish}.  While the deepest Efimov trimer is essentially unaffected by experimentally realizable confinements, we find that the first-excited trimer crosses over from a three-dimensional-like (3D-like) Efimov trimer to an extended 2D-like trimer as the attractive interactions are decreased.  This behavior is reflected in the real-space wave function for the first-excited trimer, and we furthermore quantify the three-body decay rate of the trimer by estimating the weight of the wave function at short distances. We find that this weight can be reduced by more than an order of magnitude compared with the 3D case, thus confirming our expectation that the quasi-2D geometry enhances the trimer lifetime. We discuss the optimal experimental conditions under which to realize long-lived quasi-2D trimers.

The paper is organized as follows:\,\,\,In Sec.~\ref{sec:Model_and_Methods} we outline our model of the $^{6}$Li system and our approach to determining the trimers in both three dimensions and quasi two dimensions. In Sec.~\ref{sec:trimer_states} we discuss the 3D case and then we present the quasi-2D trimer spectra and wave functions, as well as our estimate for the trimer lifetime as a function of interaction strength. We conclude in Sec.~\ref{sec:conc}.

\section{Model and Methods}
\label{sec:Model_and_Methods}

We consider three distinguishable fermions with equal masses $m$, which we label as $1$, $2$, and $3$.  If we take the case of $^6$Li atoms in the three lowest sublevels, these labels then denote the atoms' hyperfine states $|1\rangle$, $|2\rangle$, and $|3\rangle$~\cite{3-component_reference_Experimental_FeshbachRsonancesLithium_ScatteringLength_Bartenstein}.

Our goal is to model the ultracold $^6$Li system under the application of a strong transverse potential that confines the atoms to a quasi-two-dimensional geometry in the $x$-$y$ plane.  We can approximate this potential as harmonic in the $z$ direction, $V(z)=\frac12m\omega_z^2z^2$, where $\omega_z$ is the confinement frequency.  The characteristic length scale of the trap is the confinement length, $l_z=\sqrt{1/m\omega_z}$ (we work in units where $\hbar = 1$), and this always greatly exceeds the van der Waals range of the background interactions, $R_{vdW}$.  Thus, the underlying short-range interaction potential in the gas is unaffected by $V(z)$.  Before including the harmonic trap in the calculation of the trimer energies, we expound the model for the case when $\omega_z=0$ and the system is purely 3D.

\subsection{3D System}
\label{sec:3D_System}

Throughout this work we model short-range pairwise interactions that are close to resonance.  In the particular case of $^6$Li subjected to an external magnetic field $B$, there are three nearly overlapping Feshbach resonances --- one between each pair of hyperfine states --- in the range $B\sim690$ to 840 Gauss~\cite{3-component_reference_Experimental_FeshbachRsonancesLithium_ScatteringLength_Bartenstein}.  At the resonances the corresponding scattering length diverges, and all three scattering lengths remain much larger than $R_{vdW}$ throughout the range of magnetic fields considered in this study.

To model this system, we consider the Hamiltonian
\begin{align}\label{eq:Hamiltonian3d}\notag
\hat{H}=\sum_{\mathbf{q},\,i}&\,\,\epsilon_{\mathbf{q}}c^{\dagger}_{\mathbf{q},i}c_{\mathbf{q},i}+\sum_{\substack{\mathbf{q},\,\mathbf{q}',\,\mathbf{p} \\ i\,<\,j}}g_{ij}e^{-\left(q^2+q'^2\right)/\Lambda^2} \\
&\times c^{\dagger}_{\mathbf{p}/2 + \mathbf{q}',i}c^{\dagger}_{\mathbf{p}/2 - \mathbf{q}',j}c_{\mathbf{p}/2 - \mathbf{q},j}c_{\mathbf{p}/2 + \mathbf{q},i}\,,
\end{align}
where we set the volume to unity.  Here, $c^{\dagger}_{\mathbf{q},i}$ ($c_{\mathbf{q},i}$) is the second-quantized operator which creates (annihilates) an atom with 3D momentum vector $\mathbf{q}$ and label $i=1,\,2,\,3$.  The first term of the Hamiltonian corresponds to a non-interacting system where the single-particle energy is $\epsilon_\mathbf{q}=q^2/2m$ and $q\equiv|\q|$.

The second term of Eq.~\eqref{eq:Hamiltonian3d} describes the interaction of two atoms with center-of-mass momentum $\mathbf{p}$, and relative momenta $\mathbf{q}$ and $\mathbf{q}'$ before and after the collision, respectively.  To characterize the interactions between atoms $i$ and $j$, we use a separable potential of strength $g_{ij}$ with a Gaussian cutoff at a characteristic momentum $\Lambda$ \footnote{We take the cutoff $\Lambda$ to be the same for all three pairs \cite{Efimov_Lithium6_Theory_Ueda}.  This is reasonable when the van der Waals ranges of the interactions are all similar, as in the case of $^6$Li.}.  We can relate these two parameters of the model to the physical parameter of low-energy collisions, the $s$-wave scattering length $a_{ij}$, via the process of renormalization.  This results in~\cite{Werner2009,Mora2011}
\begin{align}
a_{ij}=\left(\frac{4\pi}{mg_{ij}}+\frac{\Lambda}{\sqrt{2\pi}}\right)^{-1}.\label{eq:regularise}
\end{align}
Note that, by construction, $g_{ij}= g_{ji}$ and $a_{ij}= a_{ji}$.
The renormalization condition~\eqref{eq:regularise} also allows us to determine the energy $E_2<0$ of a two-body bound state (dimer).  This is found as the pole of the $T$ matrix~\cite{Werner2009}:
\begin{alignat}{2} \label{eq:Tmatrix3D}
T_{ij}^{-1}\left(E_2\right)&=&&\,\,\frac{1}{g_{ij}}-\sum_{\q}\frac{e^{-2q^2/\Lambda^2}}{E_2-2\epsilon_{\q}}\nonumber\\&=&&\,\,\frac{m}{4\pi a_{ij}}-\frac{m\Lambda}{4\pi} F\left(mE_2/\Lambda^2\right),
\end{alignat}
where
\begin{align} \label{eq:3DFFunc}
F\left(x\right)=e^{2|x|}\sqrt{|x|}\,\mathrm{erfc}\left(\sqrt{2|x|}\,\right),
\end{align}
and $\mathrm{erfc}(x)$ is the complementary error function. Close to resonance where $a_{ij}\gg\Lambda^{-1}$, the pole condition reduces to the universal two-body energy $E_2=-1/ma_{ij}^2$.

Apart from the $s$-wave scattering length, a full description of Efimov trimers requires an additional high-energy length scale called the three-body parameter~\cite{ScatteringConcepts_NaturalLength_Universality_Efimov_Brateen}, since the trimer spectrum is unbounded from below in the absence of a short-distance cutoff.  In the model described by Eq.~\eqref{eq:Hamiltonian3d}, the three-body parameter is directly related to $\Lambda$, which thus determines the size of the deepest Efimov trimer.

We proceed now to consider the problem of three distinguishable fermions in a 3D system, and we write down a general wave function in the center-of-mass frame,
\begin{align} \label{eq:5}
|\psi_{3D}\rangle=\sum_{\q_1,\,\q_2,\,\q_3}\beta_{\q_1\q_2\q_3}|\q_1,\q_2,\q_3\rangle\,,
\end{align}
where  the state $|\q_1,\q_2,\q_3\rangle\equiv c_{\q_1,1}^{\dagger}c_{\q_2,2}^{\dagger}c_{\q_3,3}^{\dagger}|0\rangle$
and the amplitude $\beta_{\q_1\q_2\q_3}=\delta_{\q_1+\q_2+\q_3}\langle\q_1,\q_2,\q_3|\psi_{3D}\rangle$.  Projecting the Schr\"odinger equation, $\hat{H}|\psi_{3D}\rangle=E_3|\psi_{3D}\rangle$, onto an arbitrary state $\langle\q_1,\q_2,\q_3|$ then yields an expression for the three-body energy $E_{3}$:
\begin{multline}
\left(E_3-\epsilon_{\q_1}-\epsilon_{\q_2}-\epsilon_{\q_3}\right)\beta_{\q_1\q_2\q_3}= \\
 \label{eq:6}
\delta_{\q_1+\q_2+\q_3}\sum_{\{i,\,j,\,k\}} e^{-\frac{|\q_i-\q_j|^2}{4\Lambda^2}} \eta_{\q_k}^{(k)}
\,.
\end{multline}
Here, we have defined the three independent functions:
\begin{align} \label{eq:7}
\eta_{\q_i}^{(i)}=g_{jk}\sum_{\q_j,\,\q_k}\beta_{\q_1\q_2\q_3}e^{-\frac{|\q_j-\q_k|^2}{4\Lambda^2}}\,,
\end{align}
and we have $\{i,\,j,\,k\}=\{1,\,2,\,3\}$ and cyclic permutations thereof.  Rewriting the amplitudes $\beta_{\q_1\q_2\q_3}$ using Eq.~\eqref{eq:7}, we obtain three coupled expressions from Eq.~\eqref{eq:6}:
\begin{multline} \label{eq:8}
T_{ij}^{-1}\left(E_3-\frac{3}{2}\epsilon_{\q}\right)\eta_{\q}^{(k)}=\\\sum_{\q'}\frac{e^{-|\q+\q'/2|^2/\Lambda^2}e^{-|\q/2+\q'|^2/\Lambda^2}}{E_3-\epsilon_{\q}-\epsilon_{\q+\q'}-\epsilon_{\q'}}\left[\eta_{\q'}^{(i)}+\eta_{\q'}^{(j)}\right],
\end{multline}
where $\{i,\,j,\,k\}$ take the same values as above.

In the following,
we approximate the Gaussian cutoff functions appearing in Eq.~\eqref{eq:8} as
\begin{align} \label{eq:10}
e^{-|\q+\q'/2|^2/\Lambda^2}e^{-|\q/2+\q'|^2/\Lambda^2}\simeq e^{-q^2/\Lambda^2}e^{-q'^2/\Lambda^2}.
\end{align}
We have checked the validity of this step by evaluating the spectrum with and without the approximation for the case of three identical bosons.  In this case, the relative error on the three-body parameter is about $1\%$, and the relative error on the trimer energy at unitarity is similar.  We therefore expect the relative error in the $^6$Li scenario to remain very small, as well.  In particular, we expect the error to be further reduced towards the 2D regime (for large $B$ fields), where the trimer has less weight at short range~\cite{Quasi2D_EfimovTrimers_Parish}.

By relating the $T$ matrix to the scattering length $a_{ij}$ via Eq.~\eqref{eq:Tmatrix3D}, we find that $E_{3}$ satisfies a matrix equation:
\begin{align} \label{eq:11}
\begin{pmatrix}
\frac{1}{a_{12}} & 0 & 0 \\
0 & \frac{1}{a_{23}} & 0 \\
0 & 0 & \frac{1}{a_{31}} \\
\end{pmatrix}
\begin{pmatrix}
\eta^{(3)} \\
\eta^{(1)} \\
\eta^{(2)} \\
\end{pmatrix}
=
\begin{pmatrix}
D & M & M \\
M & D & M \\
M & M & D \\
\end{pmatrix}
\begin{pmatrix}
\eta^{(3)} \\
\eta^{(1)} \\
\eta^{(2)} \\
\end{pmatrix},
\end{align}
where $\eta^{(i)}$ is a column vector with elements $\eta_{q}^{(i)}$. Above, $D$ is a diagonal matrix in momentum $q$ with elements
\begin{align} \label{eq:12}
D_{q}=\Lambda F\left(\frac{mE_3}{\Lambda^2}-\frac{3q^2}{4\Lambda^2}\right),
\end{align}
while the $q^{\mathrm{th}}$ element of the matrix multiplication of $M$ onto the vector $\eta^{(i)}$ is 
\begin{align} \label{eq:13}
\left[M\eta^{(i)}\right]_q=&\,\,\int_{0}^{\infty}\frac{ q'dq'}{\pi q}e^{-\left(q^{2}+q'^{2}\right)/\Lambda^{2}}\nonumber\\&\,\times\mathrm{ln}\left[\frac{E_{3}-(q^{2}-qq'+q'^{2})/m}{E_{3}-(q^{2}+qq'+q'^{2})/m}\right]\eta_{q'}^{(i)}\,.
\end{align}
Since we are looking for bound states, we consider the $s$-wave channel and assume $E_{3}<0$.  Equations~\eqref{eq:11}--\eqref{eq:13} can be solved numerically for $E_{3}$, and the
solution for the $^6$Li system is discussed in Sec.~\ref{sec:trimer_states}.

For the case of SU(3)-symmetric interactions where $a_{12}=a_{23}=a_{31}\equiv a$, the ground state of our system reduces to that of three identical bosons. Indeed, by defining $\bar{\eta}=\eta^{(1)}+\eta^{(2)}+\eta^{(3)}$, Eq.~\eqref{eq:11} becomes
\begin{align} \label{eq:3Bosons}
\frac{1}{a}\bar{\eta}=D\bar{\eta}+2M\bar{\eta}\,,
\end{align}
which is exactly the equation for three identical bosons.

\subsection{Quasi-2D System}
\label{sec:Quasi-2D_System}

We move on to consider the scenario where the atoms are tightly harmonically confined along the $z$ direction.  In the absence of interactions, the particles occupy the ground state of the harmonic-oscillator potential and the gas is kinematically 2D. On the other hand, interacting fermions can explore all excited levels of the trap~\cite{Q2DBackground_Review_Meera}.  We thus write down a Hamiltonian where the sums run over not only each atom's momentum $\q$ (which is now an in-plane vector perpendicular to $z$), but also its harmonic-oscillator index $n$:
\begin{widetext}
\begin{align} \label{eq:q2DHam}
\hat{H}_{q2D}=\sum_{\q,\,n,\,i}\epsilon_{\q n}c_{\q n,i}^{\dagger}c_{\q n,i}+\sum_{\substack{\q,\,\q',\,\mathbf{p}\\i\,<\,j}}\sum_{\substack{n_i,\,n_j,\,n_i',\,n_j'\\N,\,n_{ij},\,n_{ij}'}}&\,\,g_{ij}e^{-\left(q^2+q'^2\right)/\Lambda^2}f_{n_{ij}}f_{n_{ij}'}\langle n_i',n_j'|N,n_{ij}'\rangle\langle N,n_{ij}|n_i,n_j\rangle\nonumber\\&\times c_{\mathbf{p}/2+\q',n_i',i}^{\dagger}c_{\mathbf{p}/2-\q',n_j',j}^{\dagger}c_{\mathbf{p}/2-\q,n_j,j}c_{\mathbf{p}/2+\q,n_i,i}\,,
\end{align}
\end{widetext}
where the operator $c^\dagger_{\q n,i}$ creates atom $i$ with in-plane momentum $\q$ and harmonic-oscillator quantum number $n\geq0$.

The first term in the Hamiltonian~\eqref{eq:q2DHam} corresponds to a non-interacting system where each particle has energy,
\begin{align} \label{eq:q2DSingleParticleEnergy}
\epsilon_{\q n}\equiv \epsilon_{\q}+\left(n+\frac12\right)\omega_z\,,
\end{align}
where, in a slight abuse of notation, we have now defined $\epsilon_\q=q^2/2m$ in terms of the in-plane momentum.
The second term of Eq.~\eqref{eq:q2DHam} accounts for the pairwise interactions.  Since these only depend on the relative motion, we model excitations in the harmonic-oscillator space by transforming from the individual quantum numbers, $n_i$ and $n_j$, to the center-of-mass $N$ and relative $n_{ij}$ quantum numbers.  Hence, the object $\langle N,n_{ij}|n_i,n_j\rangle$ in Eq.~\eqref{eq:q2DHam} is the two-body Clebsch--Gordan coefficient with the selection rule, $n_i+n_j=N+n_{ij}$~\cite{SMIRNOV1962346}.
We have also defined
\begin{align}
f_n=\sum_{q_z}\widetilde{\phi}_n\left(q_z\right)e^{-q_z^2/\Lambda^2},
\end{align}
where $\widetilde{\phi}_n(q_z)$ is the Fourier transform of the harmonic-oscillator eigenfunction. It can be shown that
\begin{align} \label{eq:q2DHOEigenFunc}
f_{2n}=\left(-1\right)^n\frac{1}{\left(2\pi l_z^2\right)^{1/4}}\frac{\sqrt{\left(2n\right)!}}{2^nn!}\frac{1}{\sqrt{1+\lambda}}\left(\frac{1-\lambda}{1+\lambda}\right)^n,
\end{align}
while $f_{2n+1}=0$~\cite{Quasi2D_EfimovTrimers_Parish}.  Above, $\lambda\equiv\left(\Lambda l_z\right)^{-2}$ is the (squared) ratio between the length scale of the short-distance physics $\Lambda^{-1}$ and the confinement length $l_z$. This ratio is very small in typical experiments~\cite{Q2DBackground_Review_Meera}.  

Under a quasi-2D confinement, the threshold energy for free-atom motion is increased since the zero-point energy of the trap must be taken into account.  The dimer energy, $E_2<\omega_z/2$, is again given by the pole of the relevant $T$ matrix, $\mathcal T$~\cite{Q2DBackground_Review_Meera}:
\begin{alignat}{2} \label{eq:q2DTMatrix}
f_0^{2}\,\mathcal{T}_{ij}^{-1}\left(E_2\right)&=&&\,\,\frac{1}{g_{ij}}-\sum_{\q,\,n}\frac{e^{-2q^{2}/\Lambda^{2}}f_{n}^{2}}{E_2-2\epsilon_{\q}-\left(n+1/2\right)\omega_z}\nonumber\\&=&&\,\,\frac{m}{4\pi l_z}\left[\frac{l_z}{a_{ij}}-\mathcal{F}\left(E_2/\omega_z-1/2\right)\right],
\end{alignat}
where
\begin{align} \label{eq:q2DFFunc}
\mathcal{F}\left(x\right)=&\int_0^\infty\frac{du}{\sqrt{4\pi(u+2\lambda)^3}}\nonumber\\&\times\left[1-\frac{\sqrt{2u+4\lambda}\,e^{xu}}{\sqrt{(1+\lambda)^2-(1-\lambda)^2\,e^{-2u}}}\right].
\end{align}
Here, because the atoms are moving in the 2D plane between interactions, we give the $\mathcal{T}$ matrix for the case where incoming and outgoing particles are in the lowest harmonic-oscillator state of the relative motion.  Furthermore, since we are considering the bare interaction in 3D (see the beginning of Sec.~\ref{sec:Model_and_Methods}), we renormalize the $\mathcal{T}$ matrix by using Eq.~\eqref{eq:regularise}.

For $l_z/a_{ij}<0$, the two-body system becomes increasingly 2D like as $|l_z/a_{ij}|$ increases~\cite{Q2DBackground_Review_Meera}.  Assuming that $\Lambda^{-1}\ll l_z$, we expand in the limit of tight confinement to obtain the expression \cite{kirk2017},
\begin{align} \label{eq:E2}
E_2\simeq-\frac{B\omega_z}{\pi} e^{-2\sqrt{\lambda}}e^{\sqrt{2\pi}\,l_z/a_{ij}}+\frac{\omega_z}2\,,
\end{align}
with $B\simeq0.905$~\cite{PhysRevA.64.012706}.  When $\lambda \to 0$ we recover the usual expression for the two-body energy.

Progressing from the two-atom system, we now consider three distinguishable fermions interacting within a quasi-2D geometry. The derivation of the three-body equation proceeds similarly to the 3D case above, so we relegate the details to the appendix, while we discuss the final solution here.  Thus, we find that the three-body energy, $E_3$, satisfies the equation:
\begin{align} \label{eq:q2DMatEq}
\begin{pmatrix}
\frac{l_z}{a_{12}} & 0 & 0 \\
0 & \frac{l_z}{a_{23}} & 0 \\
0 & 0 & \frac{l_z}{a_{31}} \\
\end{pmatrix}
\begin{pmatrix}
\eta^{(3)} \\
\eta^{(1)} \\
\eta^{(2)} \\
\end{pmatrix}
=
\begin{pmatrix}
\mathcal{D} & \mathcal{M} & \mathcal{M} \\
\mathcal{M} & \mathcal{D} & \mathcal{M} \\
\mathcal{M} & \mathcal{M} & \mathcal{D} \\
\end{pmatrix}
\begin{pmatrix}
\eta^{(3)} \\
\eta^{(1)} \\
\eta^{(2)} \\
\end{pmatrix},
\end{align}
where $\eta^{(i)}$ is a tensor in momentum $q$ and harmonic-oscillator quantum number $n$ with elements $\eta_{q,n}^{(i)}$. Likewise, $\mathcal{D}$ is a diagonal tensor
\begin{align} \label{eq:q2DMatD}
\mathcal{D}_{q,n}=\mathcal{F}\left[\frac{E_3-(3q^2)/(4m)-(n+1)\omega_z}{\omega_z}\right],
\end{align}
while the matrix multiplication of the tensors $\mathcal{M}$ and $\eta$ gives
\begin{align} \label{eq:q2DMatM}
&\left[\mathcal{M}\eta^{(i)}\right]_{q,n}=-\frac{2l_z}{m}\int_{0}^{\infty}q'dq'\,\times\nonumber\\&\sum_{\substack{l,\,l'\\n'}}\frac{e^{-(q^{2}+q'^{2})/\Lambda^{2}}f_lf_{l'}\langle n,l|n',l'\rangle\eta_{q',n'}^{(i)}}{\sqrt{\left[E_3-\frac{q^2+q'^2}{m}-(n+l+1)\omega_z\right]^2-\left(\frac{qq'}{m}\right)^2}}\,,
\end{align}
where, as in the 3D problem, we have projected onto the $s$-wave sector.  Here, $q$ is the relative momentum in the $x$-$y$ plane between two atoms' center of mass and the third particle, while $l$ and $n$ are the harmonic-oscillator indices that correspond, respectively, to relative atom-atom and atom-pair motion in the $z$ direction.  Due to the raised three-body continuum, we now have $E_{3}<\omega_z$ (note that we have removed the zero-point motion corresponding to the center-of-mass motion).

In Eq.~\eqref{eq:q2DMatM}, the scalar quantity $\langle n,l|n',l'\rangle$ is the atom-pair Clebsch--Gordan coefficient where the quantum numbers satisfy $n+l=n'+l'$.  To evaluate these, we can exploit~\cite{Quasi2D_EfimovTrimers_Parish,PhysRevA96032701} their relation to Wigner's $d$ matrix~\cite{Wigner1959}:
\begin{align} \label{eq:q2DWignerD}
\langle n,l|n',l'\rangle=d_{\frac{n'-l'}2,\frac{n-l}2}^{\frac{n+l}2}\left(4\pi/3\right).
\end{align}
Similar to the situation in 3D where there are only trimers in the $s$-wave channel, here the odd and even atom-pair motions decouple and we find bound states in the even-$n$ channel only. We solve Eqs.~\eqref{eq:q2DMatEq}--\eqref{eq:q2DMatM} numerically for the $^6$Li system and discuss our results in the ensuing section.

\section{Trimer States in $^6$Li}
\label{sec:trimer_states}

\begin{figure}[t]
\centering
\includegraphics[width=\columnwidth]{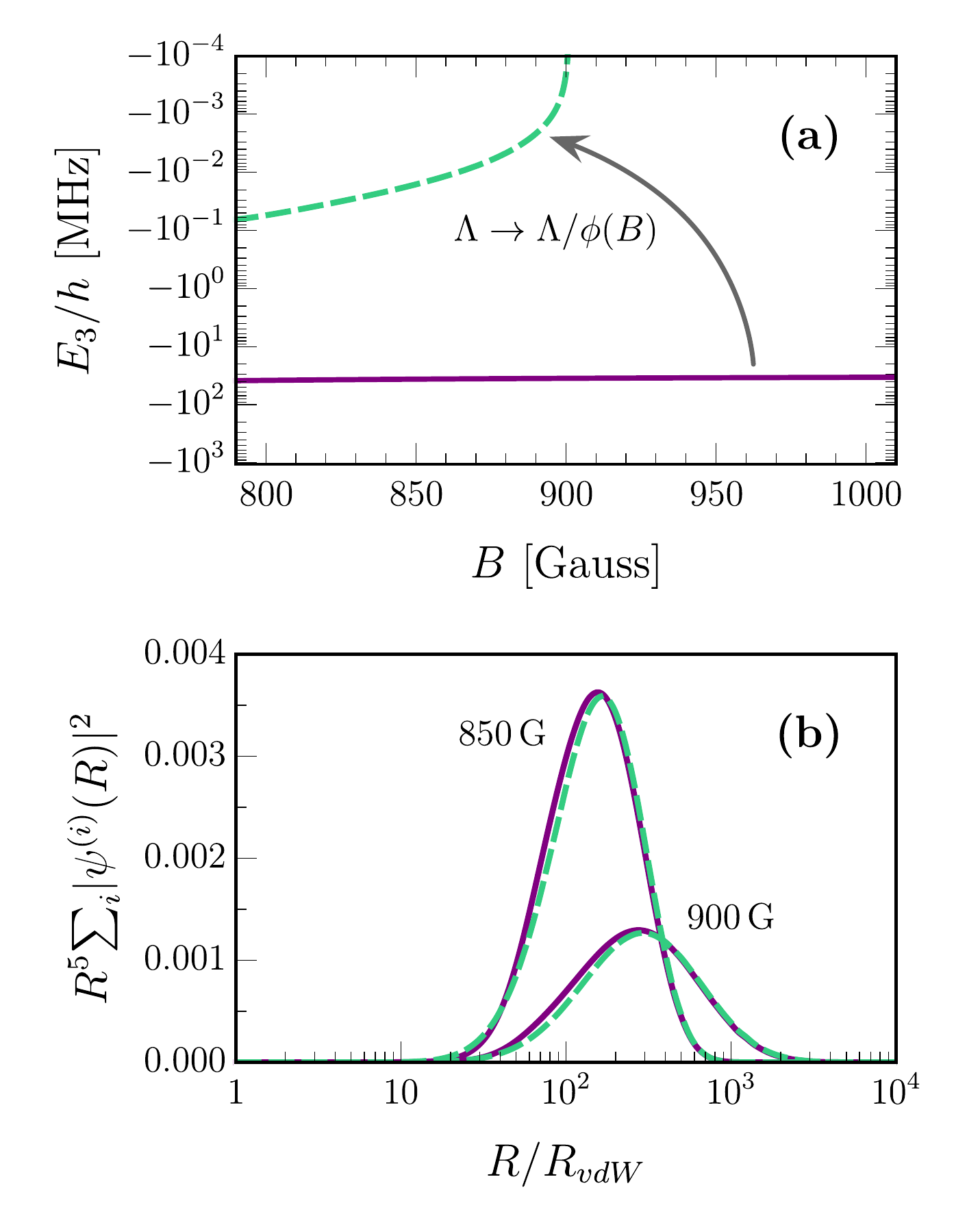}
\caption{(a)\,\,\,Spectrum of 3D trimers and illustration of rescaling:\,\,\,By modifying the momentum cutoff, $\Lambda\to\Lambda/\phi(B)$, we map the deeply bound ground-state trimer [purple solid line] in the 3D system to the excited trimer state [green dashed line].  This effectively removes the original ground state from the problem without changing the low-energy physics.\,\,\,(b)\,\,\,Probability densities (see Sec.~\ref{sec:Wave_Functions}) for the excited 3D trimer without rescaling [green dashed line] and the ground-state trimer with rescaling [purple solid line], at different magnetic fields. The functions plotted are normalized to 1.}
\label{fig:3DTrimers}
\end{figure}

\begin{figure*}[t]
\centering
\noindent\makebox[\textwidth]{%
\includegraphics[width=1.15\textwidth]{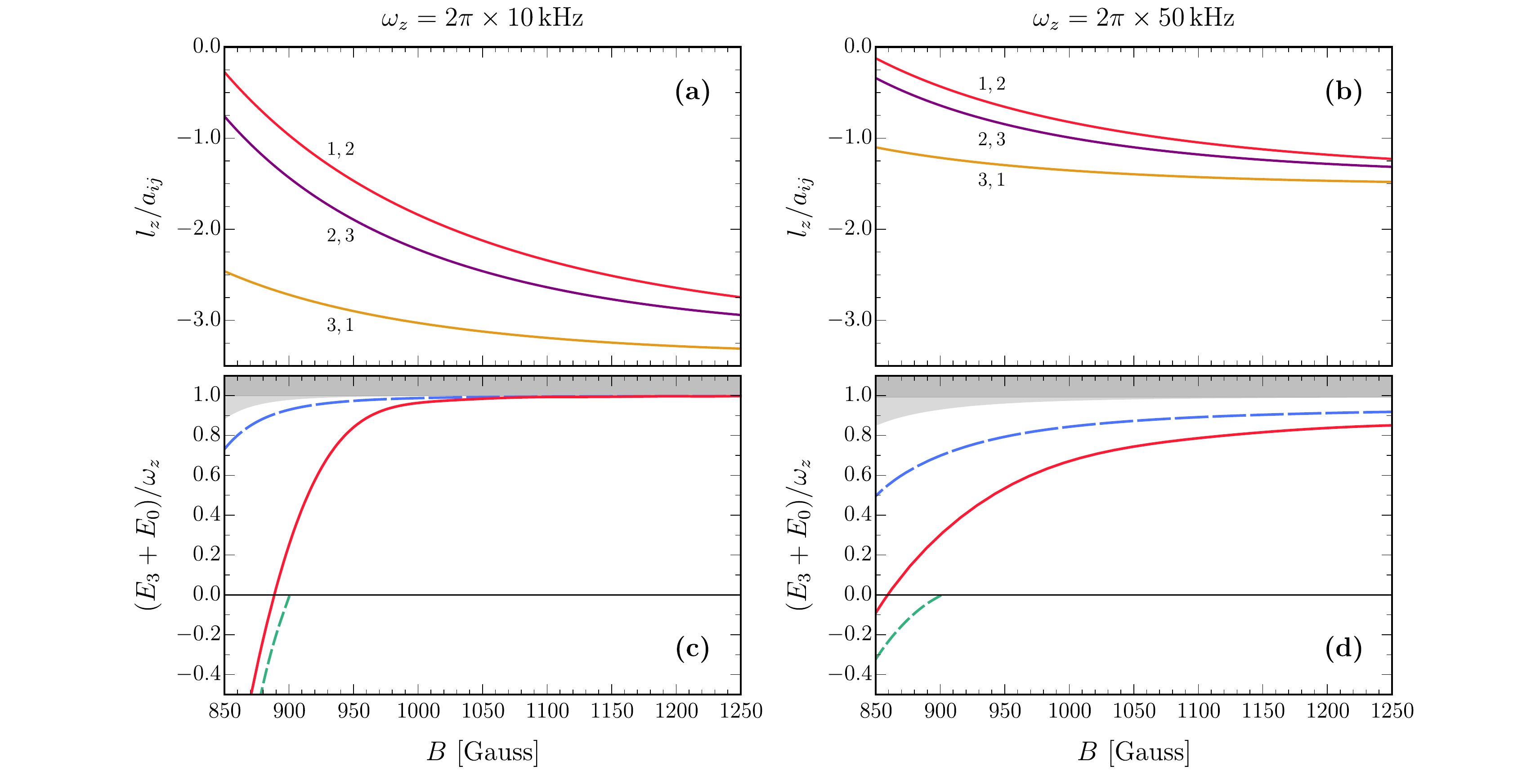}}
\caption{Energy spectra of trimers comprising atoms in the three lowest hyperfine states of $^{6}$Li, for two different confinement strengths, $\omega_z=2\pi\times10\,$kHz [left] and $\omega_z=2\pi\times50\,$kHz [right].  Panels (a) and (b) show the ratio of the confinement length $l_z$ to the $s$-wave scattering length $a_{ij}$ for all three pairs of atoms $i$ and $j$.  In panels (c) and (d) we plot the 3D, quasi-2D, and 2D trimer states as the green short-dashed, red solid, and blue long-dashed lines, respectively.  Under confinement, the threshold energy of the free-atom continuum is increased from $E_0=0$ in the 3D system to $E_0=\omega_z$ in the 2D and quasi-2D cases, which is indicated by the darker rendered area.  The lighter rendered area is the atom-dimer continuum of the confined systems, and corresponds to atoms 1 and 2 forming a dimer.}
\label{TrimerEnergiesFigure}
\end{figure*}

We now investigate the spectrum of trimers formed from atoms in the three lowest hyperfine states of $^6$Li.
Before proceeding, we discuss the scattering parameters appearing in the Hamiltonian~\eqref{eq:Hamiltonian3d} for this specific system.  The scattering lengths are obtained by using the formula~\cite{3-component_reference_Experimental_FeshbachRsonancesLithium_ScatteringLength_Bartenstein},
\begin{align}
a_{ij}=a_{bg}^{(ij)}[1+\Delta^{(ij)}(B-B_0^{(ij)})^{-1}][1+\alpha^{(ij)}(B-B_0^{(ij)})]\,.
\end{align}
Here, $a_{bg}^{(ij)}$ is the background scattering length, $\Delta^{(ij)}$ is the resonance width, $B_0^{(ij)}$ is the position of the resonance, and $\alpha^{(ij)}$ is a correction parameter~\footnote{For completeness, the respective values of $a_{bg}$, $\Delta$, $B_0$, and $\alpha$ are\,\,\,$-1405\,a_0$, $300\,$G, $834.149\,$G, and $0.0004\,$G$^{-1}$ for channel $(1,2)$;\,\,\,$-1727\,a_0$, $122.3\,$G, $690.43\,$G, and $0.0002\,$G$^{-1}$ for channel $(1,3)$;\,\,\,and $-1490\,a_0$, $222.3\,$G, $811.22\,$G, and $0.000395\,$G$^{-1}$ for channel $(2,3)$~\cite{3-component_reference_Experimental_FeshbachRsonancesLithium_ScatteringLength_Bartenstein}.}.  The relative error in this expression is expected to be less than $1\%$ over the range of magnetic fields between $B=600$ and $1200$ Gauss~\cite{3-component_reference_Experimental_FeshbachRsonancesLithium_ScatteringLength_Bartenstein}.

The second parameter of the model is the three-body parameter, i.e., the short-range length scale which ensures that the ground-state energy is well defined. The three-body parameter for $^6$Li has been calculated by using numerous models and methods (see pp.~43~and~44 of Ref.~\cite{ShimpeiReview} for a recent summary). For large fields $B>600\,$G, the three-body parameter can be calculated by fitting to experimentally measured loss rates~\cite{3-component_Efimov_RecombinationLoss2_Hara,PhysRevA.90.043636,3-component_BindingEnergyMeasurement_Ueda} and by radio-frequency spectroscopy~\cite{3-component_BindingEnergyMeasurement_Ueda,FequencyAssociation_EfimovTriemrs_jochim}. The values reported vary by roughly $10\%$ for magnetic fields ranging from~$685$~to~$895\,$G~\cite{3-component_BindingEnergyMeasurement_Ueda,PhysRevA.90.043636}. In this work, we apply a Gaussian cutoff characterized by the ultraviolet momentum scale, $\Lambda^{-1}\simeq0.89\,R_{vdW}$~\cite{Efimov_Lithium6_Theory_Ueda} where $R_{vdW}\simeq31\,a_0$~\cite{Chin821225} ($a_0$ is the Bohr radius). Such a model has been applied in previous studies~\cite{3-component_BindingEnergyMeasurement_Ueda,Efimov_Lithium6_Theory_Ueda} to fit the loss rate associated with the excited trimer crossing into the three-atom continuum, as measured in Ref.~\cite{3-component_Efimov_RecombinationLoss2_Hara}. This feature is within the range of magnetic fields that we consider.

In the following, we present results for two confinement strengths, $\omega_z=2\pi\times10\,$kHz and $\omega_z=2\pi\times50\,$kHz, corresponding to confinement lengths of $7800\,a_0$ and $3500\,a_0$, respectively.  Both of these are within reach of current experiments on quasi-2D two-component $^6$Li gases~\cite{From2Dto3D_Experimental_Vale,3-component_reference_Experimental_Quasi2D_Pairing_Experimental_Jochim,2Dexperiment_tightconfinement_Lithium6_Zwierlein,PhysRevLett.117.093601}.

\subsection{Trimer Energies}
\label{sec:Trimer_Energies}

In three dimensions, it has been predicted that there exist two trimers~\cite{3-component_Efimov_RecombinationLoss2_Hara}.  Solving Eq.~\eqref{eq:11} for the trimer energies, we find the spectrum shown in Fig.~\ref{fig:3DTrimers}(a), which agrees with the results of Ref.~\cite{Efimov_AtomDimer_Resonance_Ueda}.  Notice how the excited trimer only exists for magnetic fields $\lesssim900\,$G, beyond which it disappears into the three-atom continuum.  On the other hand, the ground-state trimer is expected to be very deeply bound, with a binding energy $\sim2\pi\times30\,$MHz that remains relatively constant over the range of magnetic fields investigated~\footnote{The ground-state trimer can furthermore be strongly affected by non-universal effects beyond our theory~\cite{Efimov_AtomDimer_Resonance_Ueda}. However, its binding energy is still $\sim2\pi\times10\,$MHz which greatly exceeds experimentally realistic confinement strengths.}.

The large separation of energy scales in the 3D trimer spectrum, Fig.~\ref{fig:3DTrimers}(a), presents a significant challenge to calculating the spectrum in the presence of confinement.  In particular, the deepest trimer energy exceeds realistic confinement strengths by three orders of magnitude, and thus we may expect this state to be essentially unaffected by the confinement. This, in turn, means that the number of harmonic-oscillator levels taken into account in the numerics has to greatly exceed 1000 to properly describe all energy scales of the problem --- which is in practice unfeasible.  Instead, we take advantage of the fact that we are primarily interested in the excited trimer at low energies.  
Therefore we can rescale the cutoff, $\Lambda\to\Lambda/\phi(B)$, in such a way that the ground state of the rescaled model coincides with the excited state of the original model --- see Fig.~\ref{fig:3DTrimers}(a).
This procedure effectively removes the ground state of the original problem. 
The rescaling of the short-range parameter is inspired by the system of three identical bosons, where the spectrum at large scattering length is characterized by a discrete scaling symmetry, such that the low-energy physics is unchanged under a rescaling of the ultraviolet cutoff:\,\,\,$\Lambda\to \Lambda/22.7$. 

The method described above yields a scaling function $\phi(B)$ which decreases approximately linearly with increasing magnetic-field strength, i.e., from $\phi(B=840\,\rm{G})\simeq20.9$ to $\phi(B=900\,\rm{G})\simeq18.7$.  However, we cannot compute the scaling parameter 
for fields $B\gtrsim 900\,$G, since there the excited 3D trimer ceases to exist.
Hence, we simply take $\phi(B=900\,\text{G})$ throughout this regime, since the three-body parameter we use is most accurate at $\sim900\,$G, where the excited trimer disappears and a loss feature is observed~\cite{Efimov_Lithium6_Theory_Ueda}.

In Fig.~\ref{fig:3DTrimers}(b) we show how, outside the short-range region, the wave function (see Section~\ref{sec:Wave_Functions}) of the excited state in the original model closely matches that of the ground state with the rescaled cutoff. This result corroborates the use of our rescaling.
While such an approach introduces effective range corrections to our quasi-2D results [see Eq.~\eqref{eq:E2}], these are expected to be small in the experimental regime of interest since the rescaled van der Waals range remains much smaller than the confinement length.

In Fig.~\ref{TrimerEnergiesFigure}, we present our calculated trimer energies for the case of a quasi-2D geometry with confinement strengths, $\omega_z=2\pi\times10\,$kHz and $\omega_z=2\pi\times50\,$kHz. As discussed above, we only show the excited trimer.  For both confinement strengths, we see how the trimer energy is close to that of the 3D trimer for $B\lesssim900\,$G.  At larger magnetic fields, the trimer is stabilized by the confinement and exists far beyond its regime of existence in three dimensions. In particular, we see that the binding energy of the trimer can be comparable to $\omega_z$ for a large range of magnetic fields beyond $900\,$G.  The existence of the trimer in this regime may be understood from how the three-body continuum in quasi-2D is raised by $\omega_z$, which results in an effective long-range attractive well in the hyperspherical potential~\cite{Quasi2D_EfimovTrimers_Parish}.  Indeed, this result is analogous to how the two-body state is stabilized by a confining potential~\cite{PhysRevA.64.012706}.

We may elucidate our results further by considering the 2D limit.  When all three scattering lengths are negative and their magnitudes are less than the confinement length, the few-body states are expected to be extended in the plane and thus strongly modified from their three-dimensional counterparts.  As shown in Figs.~\ref{TrimerEnergiesFigure}(a)~and~\ref{TrimerEnergiesFigure}(b), for a confinement of $\omega_z=2\pi\times10\,$kHz this condition is satisfied when $B\gtrsim900\,$G, while for $\omega_z=2\pi\times50\,$kHz the 2D condition requires stronger magnetic fields, $B\gtrsim1100\,$G.  In this regime of large $B$ fields, the trimer energies are expected to approach those predicted from a purely 2D theory.  We obtain the 2D limit by taking just one atom-pair harmonic-oscillator state in the three-body equation~\eqref{eq:q2DMatEq}, while still retaining the full quasi-2D $\mathcal{T}$ matrix [i.e., the exact ${\cal D}$ in Eq.~\eqref{eq:q2DMatD}], since this allows us to accommodate any effective range that arises from the confinement and acts through the two-body physics.  Indeed, in Figs.~\ref{TrimerEnergiesFigure}(c)~and~\ref{TrimerEnergiesFigure}(d) we see that the quasi-2D trimer approaches the 2D limit at large magnetic fields.

We also note how, in the case where the scattering lengths are equal, it is predicted that two trimers exist in the 2D limit~\cite{bruch1979binding}.  For $^6$Li, the ratios between the three scattering lengths approach unity for increasingly strong magnetic fields.  Therefore, eventually one would expect a second quasi-2D trimer to emerge from the continuum in this regime.  However, for the confinement strengths considered here, the second 2D trimer remains very weakly bound on the scale shown in the figure.

\begin{figure*}[t]
\centering
\noindent\makebox[\textwidth]{%
\includegraphics[width=1.05\textwidth,left]{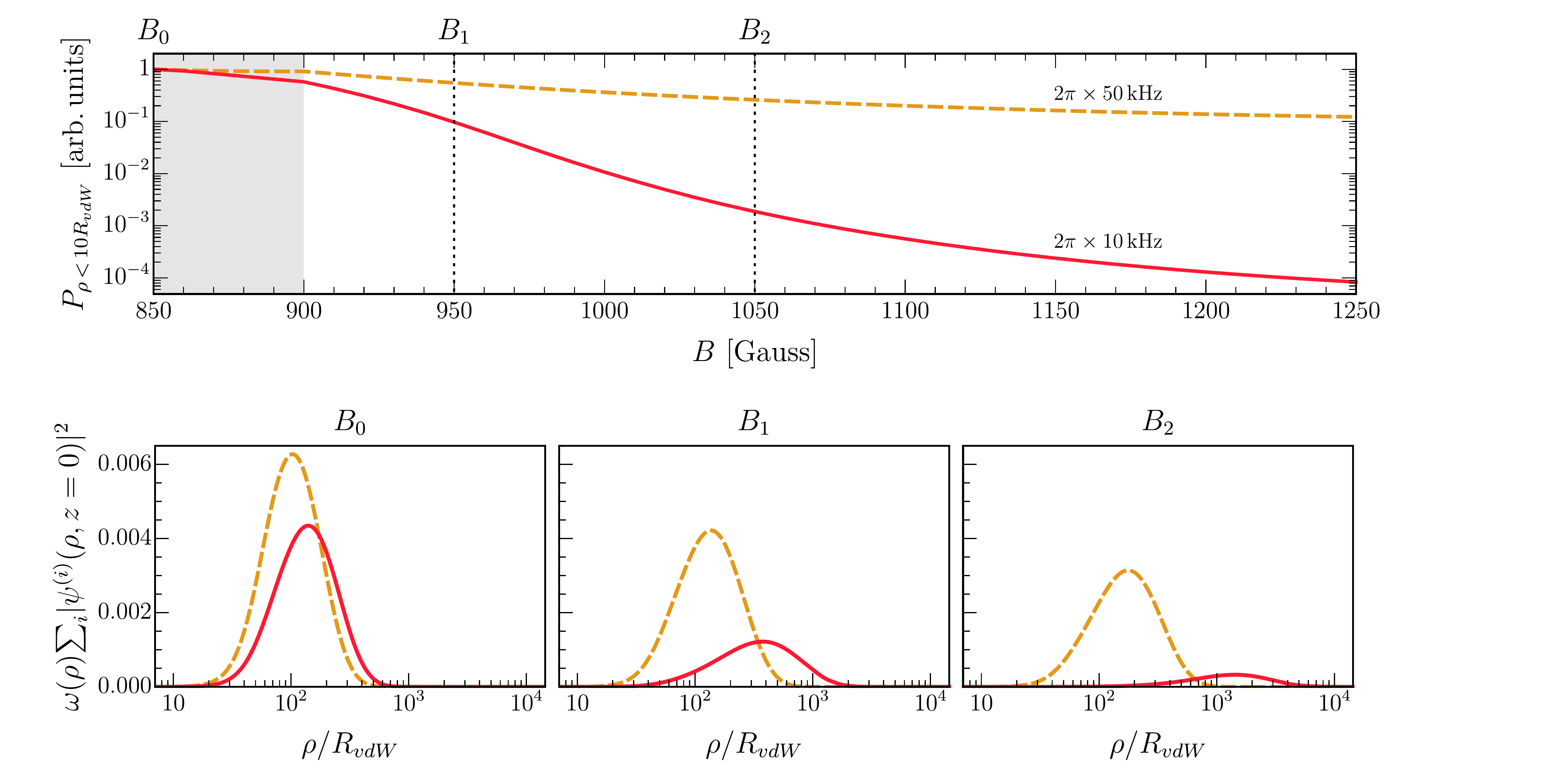}}
\caption{(upper panel)\,\,\,The relative weight of the trimer wave function at short range (\,$\lesssim10\,R_{vdW}$) in a quasi-2D geometry for two different confinement strengths, $\omega_z=2\pi\times10$\,kHz [red solid line] and $\omega_z=2\pi\times50$\,kHz [orange dashed line].  The shaded area indicates the regime of existence for the excited trimer in 3D.  Note that the small kink in the relative weight at $B=900\,$G is related to our rescaling and should not be understood as a physical effect.\,\,\,(lower panels)\,\,\,The in-plane quasi-2D probability densities for both confinements at three different magnetic-field strengths.  [Refer to the discussion around Eqs.~\eqref{eq:weight}~and~\eqref{eq:pq2D} of the text for our treatment of the quasi-2D wave functions.]  The functions plotted are normalized to 1.}
\label{fig:q2DWaveFunctions}
\end{figure*}

\subsection{Wave Functions}
\label{sec:Wave_Functions}

We now analyze how the quasi-2D confinement affects the trimer wave functions and, in particular, their relative weight at short distance. 
Starting with the 3D case, we consider the real-space \emph{atom-pair} wave function defined as the following Fourier transform:
\begin{align}\label{eq:3Dwave}
    \psi^{(i)}(R)=\frac{1}{\sqrt{\mathcal{N}_{i}}} 
    \int \frac{q\,dq}{R}\sin(qR)\,\eta^{(i)}_{q}\, ,
\end{align}
where the constant $\mathcal{N}_{i}$ ensures normalization. Here we take advantage of the fact that the trimer states satisfying Eq.~\eqref{eq:11} have $s$-wave symmetry and thus $\eta^{(i)}_{\q}$ does not depend on the direction of $\q$.  This wave function corresponds to the scenario where we take two atoms to have zero separation and then consider the motion of this pair with the remaining atom $i$.  As such, we have three atom-pair wave functions $\psi^{(i)}(R)$, one for each pair, where $R$ corresponds to the relative atom-pair coordinate. The 3D wave functions are illustrated in Fig.~\ref{fig:3DTrimers}(b).

Likewise, we define the quasi-2D real-space atom-pair wave function, from the solution of the quasi-2D three-body equation~\eqref{eq:q2DMatEq}, as
\begin{align}\label{eq:q2Dwave}
 \psi^{(i)}(\rho,z)&=\frac{1}{\sqrt{\mathcal{N}^{(\rm{q2D})}_{i}}} 
\int q\,dq\sum_{n}f_n(z)\,J_0(q\rho)\,\eta^{(i)}_{q,n} \, ,
\end{align}
where $\mathcal{N}^{(\rm{q2D})}_{i}$ is again the normalization and $J_0$ is the Bessel function. Here, $z$ is the atom-pair coordinate in the transverse direction, while $\rho$ is the separation in the plane. We show these wave functions at three different magnetic fields in Fig.~\ref{fig:q2DWaveFunctions}.

To evaluate the weight of the three-body wave functions at short distance, we employ the following approximation to convert the atom-pair wave functions, Eqs.~\eqref{eq:3Dwave}~and~\eqref{eq:q2Dwave}, to those describing the full three-particle problem:\,\,\,First, we note that in the case of identical pairwise interactions, the 3D three-atom hyperspherical wave function is approximately related to the atom-pair wave function by multiplying $\psi^{(i)}(R)$ by $R^{5/2}$, where $R$ is interpreted as the three-body hyperradius \cite{ScatteringConcepts_NaturalLength_Universality_Efimov_Brateen}. Similarly, the 2D three-atom hyperspherical wave function is obtained by multiplying $\psi^{(i)}(\rho,0)$ by $\rho^{3/2}$ \cite{nielsen1999}, where again $\rho$ corresponds to the planar hyperradius.  In the $^6$Li case, we still expect this to be a reasonable approximation since the three interaction strengths are approximately equal.  Therefore, for the quasi-2D system, we define the following weighting function that interpolates between the 2D and 3D limits:
\begin{align}
\omega(\rho)=\frac{\rho^5}{\rho^2+\frac32\,l_z^2}\,.
\label{eq:weight}
\end{align}
We then define the relative weight of the trimer at short distances as
\begin{align}
P_{\rho\,<\,\rho_0}=\frac{\int_0^{\rho_0}d\rho\,\omega(\rho)\sum_{i\,=\,1}^3|\psi^{(i)}(\rho,z=0)|^2}{\int_0^\infty d\rho\,\omega(\rho)\sum_{i\,=\,1}^3|\psi^{(i)}(\rho,z=0)|^2}\,.
\label{eq:pq2D}
\end{align}

At this stage, several comments are in order:\,\,\,First, in the following, we take the short-range length scale to be $\rho_0=10\,R_{vdW}$; we have checked that our results are not sensitive to the precise range, by varying the definition of this length scale up to a factor of 10. Second, the crossover scale of $\frac32\,l_z^2$ in Eq.~\eqref{eq:weight} is the squared atom-pair confinement length; again, we have checked that our results do not depend sensitively on the precise range chosen for this interpolation.
Third, we evaluate the weight at short range by taking $z=0$. This is reasonable when the wave function is 3D like, since it is then isotropic and we are thus free to choose any direction. Conversely, when the wave function is more 2D like at large distances, then the relevant part of the wave function is exactly the $z=0$ component.

In Fig.~\ref{fig:q2DWaveFunctions}, we show our calculated trimer weight in the short-range regime and the corresponding probability densities at select magnetic fields.  Beyond $B\simeq900$\,G, where the excited 3D trimer ceases to exist, the short-range relative weight of the quasi-2D trimer decreases by almost an order of magnitude for the stronger confinement, and four orders of magnitude for the weaker confinement, over the range of magnetic fields shown.  Thus, we expect the lifetime of the trimer to increase accordingly.  The reduction in the short-range weight is due to the trimer becoming increasingly spatially extended as we approach the 2D limit.  This is because, unlike in 3D, the trimer now resides in the long-range attractive tail of the hyperspherical potential.  This is the same mechanism responsible for the longer lifetimes of the trimers of identical bosons discussed in Ref.~\cite{Quasi2D_EfimovTrimers_Parish}.  Note that our approximate expression~\eqref{eq:pq2D} does not account for how the two-body scattering within each pair of atoms
changes from 3D to 2D.  However, if anything, we would expect 2D-like two-body scattering to further suppress decay of the trimers into atoms and dimers~\cite{3BodyRecombination_2D__Incao,ThreeBosons_2D_Hammer,ngampruetikorn2013}.

\section{Conclusions and Outlook}
\label{sec:conc}

In this work, we have considered the problem of three distinguishable fermions confined to a quasi-2D geometry.  In particular, we have allowed for the possibility that the three pairwise interactions are different, as is the case for the $^6$Li system.  While trimers comprising three dissimilar particles can, in principle, also be manufactured from bosons, we have exclusively studied the quasi-2D Fermi gas since the corresponding Bose system has significant instabilities~\cite{huang2018}.  Furthermore, the $^6$Li system has the advantage that the two-component Fermi gas is stable, and the three-component trimers in 3D have already been realized in experiment~\cite{FequencyAssociation_EfimovTriemrs_jochim}.  Thus, by using realistic experimental parameters, we have computed the $^6$Li trimer spectrum for two different quasi-2D confinements.  We have focused exclusively on the evolution of the excited trimer from the 3D spectrum, since the ground-state trimer is too deeply bound to be significantly affected by the confinement.  We have found that the excited trimer evolves into a 2D-like spatially extended trimer as the interactions are decreased with increasing magnetic field.  This behavior is also apparent in the approximate three-body wave function we have calculated for the trimer.

Our results indicate that the quasi-2D trimers can be longer lived by at least an order of magnitude compared with their 3D counterparts, since these spatially extended trimers have a reduced probability that three fermions can approach each other at short distances and decay into a deeply bound dimer state.  This opens the door to engineering long-lived three-body bound states in cold-atom experiments.  In principle, such trimers can be associated from atoms and pairs in a quasi-2D two-component $^6$Li gas. To achieve this in experiment, we require all interactions $l_z/a_{ij} < -1$  and a moderately strong quasi-2D confinement, such that the trimer is sufficiently bound and sufficiently spatially extended. If the confinement is too strong, i.e., when $\omega_z$ is around $2\pi \times$50 kHz or more, then the trimer lifetime will become comparable to that in 3D, while if the confinement is too weak, then the quasi-2D trimer will be dissociated by thermal fluctuations.  Since the temperature of the confined $^{6}$Li gas is typically of order kHz~\cite{From2Dto3D_Experimental_Vale,2Dexperiment_tightconfinement_Lithium6_Zwierlein,3-component_reference_Experimental_Quasi2D_Pairing_Experimental_Jochim,PhysRevLett.117.093601}, we expect the optimal confinement and magnetic field to be in the ranges 2$\pi\times$10--20 kHz and 950--1000 Gauss, respectively.

\acknowledgments We are grateful to C.~Vale, P.~Dyke, and S.~Hoinka for fruitful discussions.\,\,\,J.L. is supported through the Australian Research Council Future Fellowship FT160100244.\,\,\,M.M.P. and J.L. also acknowledge financial support from the Australian Research Council via Discovery Project No.~DP160102739.

\onecolumngrid

\appendix

\section*{Appendix:\,\,\,\,\,\,\,T\lowercase{hree}-B\lowercase{ody} P\lowercase{roblem in a} Q\lowercase{uasi}-2D S\lowercase{ystem}}

Here, we derive Eqs.~\eqref{eq:q2DMatEq}--\eqref{eq:q2DMatM} of the main text which determine the bound states of three distinguishable fermions interacting in a quasi-2D geometry.

We write down a general wave function at zero center-of-mass momentum,
\begin{align} \label{eq:q2DWaveF}
|\psi_{q2D}\rangle=\sum_{\substack{\q_1,\,\q_2,\,\q_3\\n_1,\,n_2,\,n_3}}\beta_{n_1n_2n_3}^{\q_1\q_2\q_3}|\q_1n_1,\q_2n_2,\q_3n_3\rangle\,,
\end{align}
where the state $|\q_1n_1,\q_2n_2,\q_3n_3\rangle\equiv c_{\q_1n_1,1}^\dagger c_{\q_2n_2,2}^\dagger c_{\q_3n_3,3}^\dagger|0\rangle$ and the amplitude $\beta_{n_1n_2n_3}^{\q_1\q_2\q_3}=\delta_{\q_1+\q_2+\q_3}\langle\q_1n_1,\q_2n_2,$ $\q_3n_3|\psi_{q2D}\rangle$.  For three particles, we transform from the individual harmonic-oscillator indices $\{n_1,\,n_2,\,n_3\}$ to the new indices $\{n_{ij},\,n_k^{ij},\,N\}$.  These correspond, respectively, to the relative motion of two atoms in the $z$ direction, $z_{ij}=z_i-z_j$, the relative motion between their center of mass and the third atom, $z_k^{ij}=(z_i+z_j)/2-z_k$, and the center-of-mass motion of all three atoms, $Z=(z_i+z_j+z_k)/3$~\cite{Bradly2014,PhysRevA96032701}.  After projecting the Schr\"odinger equation $\hat{H}_{q2D}|\psi_{q2D}\rangle=E_3|\psi_{q2D}\rangle$ onto an arbitrary state, we obtain the following expression for the three-body energy $E_3$:
\begin{align} \label{eq:q2DEquation}
\left(E_3-\epsilon_{\q_1n_1}-\epsilon_{\q_2n_2}-\epsilon_{\q_3n_3}\right)\beta_{n_1n_2n_3}^{\q_1\q_2\q_3}=&\,\sum_{\substack{\q_1',\,\q_2',\,\q_3'\\\{i,\,j,\,k\}}}\sum_{\substack{n_1',\,n_2',\,n_3'\\n_k^{ij},\,n_{ij},\,n_{ij}'}}g_{ij}e^{-\frac{|\q_i-\q_j|^2}{4\Lambda^2}}e^{-\frac{|\q_i'-\q_j'|^2}{4\Lambda^2}}\delta_{\q_k,\q_k'}\delta_{\q_1+\q_2+\q_3}f_{n_{ij}}f_{n_{ij}'}\nonumber\\&\times\langle n_1,n_2,n_3|N=0,n_k^{ij},n_{ij}\rangle\langle N=0,n_k^{ij},n_{ij}'|n_1',n_2',n_3'\rangle\beta_{n_1'n_2'n_3'}^{\q_1'\q_2'\q_3'}\,,
\end{align}
where $\epsilon_{\q n}$ is defined in Eq.~\eqref{eq:q2DSingleParticleEnergy} and we have $\{i,\,j,\,k\}=\{1,\,2,\,3\}$ and cyclic permutations.  Note that since we are working in the center-of-mass frame, we make the simplification $N=0$.

We can remove two harmonic-oscillator indices from the problem by defining three independent functions~\cite{PhysRevA96032701},
\begin{align} \label{eq:q2DEta}
\eta_{\q_k',n_k^{ij}}^{(k)}=g_{ij}\sum_{\substack{\q_1,\,\q_2,\,\q_3\\n_{ij},\,n_1,\,n_2,\,n_3}}e^{-\frac{|\q_i-\q_j|^{2}}{4\Lambda^2}}\delta_{\q_k,\q_k'}f_{n_{ij}}\langle 0,n_k^{ij},n_{ij}|n_1,n_2,n_3\rangle\beta_{n_1,n_2,n_3}^{\q_1,\q_2,\q_3}\,,
\end{align}
which allow us to rewrite Eq.~\eqref{eq:q2DEquation} as
\begin{align} \label{eq:q2DNewEqEta}
&\left(E_3-\epsilon_{\q_1n_1}-\epsilon_{\q_2n_2}-\epsilon_{\q_3n_3}\right)\beta_{n_1n_2n_3}^{\q_1\q_2\q_3}=\sum_{\substack{n_k^{ij},\,n_{ij}\\\{i,\,j,\,k\}}}e^{-\frac{|\q_i-\q_j|^2}{4\Lambda^2}}\delta_{\q_1+\q_2+\q_3}f_{n_{ij}}\langle n_1,n_2,n_3|0,n_k^{ij},n_{ij}\rangle\eta_{\q_k,n_k^{ij}}^{(k)}\,.
\end{align}
To proceed, we divide by $(E_3-\epsilon_{\q_1n_1}-\epsilon_{\q_2n_2}-\epsilon_{\q_3n_3})$, and then act with the operator
\begin{align} \label{eq:q2DOp}
g_{ij}\sum_{\substack{\q_1,\,\q_2,\,\q_3\\n_{ij}',\,n_1,\,n_2,\,n_3}}e^{-\frac{|\q_i-\q_j|^2}{4\Lambda^2}}\delta_{\q_k,\q_k'}f_{n_{ij}'}\langle 0,n_k^{ij\prime},n_{ij}'|n_1,n_2,n_3\rangle(\,\cdot\,)
\end{align}
on the left three separate times, where $\{i,\,j,\,k\}$ take the same values as in Eqs.~\eqref{eq:q2DEquation}--\eqref{eq:q2DNewEqEta}.  This yields a separate equation for each of the three $\eta^{(k)}$ functions, and we give one of these below:
\begin{align} \label{eq:q2D1stEq12}
\eta_{\q_3',n_3^{12\prime}}^{(3)}=g_{12}\sum_{\substack{\q_1,\,\q_2,\,\q_3\\n_k^{ij},\,n_{ij},\,n_{12}'\\\{i,\,j,\,k\}}}\frac{e^{-|\q_i-\q_j|^2/(4\Lambda^2)}e^{-|\q_1-\q_2|^2/(4\Lambda^2)}}{E_3-\epsilon_{\q_1}-\epsilon_{\q_2}-\epsilon_{\q_3}-\left(n_3^{12\prime}+n_{12}'+1\right)\omega_z}\delta_{\q_3,\q_3'}\delta_{\q_1+\q_2+\q_3}f_{n_{ij}}f_{n_{12}'}\nonumber\\\times\,\langle 0,n_3^{12\prime},n_{12}'|0,n_k^{ij},n_{ij}\rangle\eta_{\q_k,n_k^{ij}}^{(k)}\,.
\end{align}
To arrive at Eq.~\eqref{eq:q2D1stEq12}, we make use of the fact that
\begin{align} \label{eq:q2DIdentity}
\sum_{n_1,\,n_2,\,n_3}|n_1,n_2,n_3\rangle\frac{1}{E_3-\epsilon_{\q_1n_1}-\epsilon_{\q_2n_2}-\epsilon_{\q_3n_3}}\langle n_1,n_2,n_3|=\frac{1}{E_3-\epsilon_{\q_1}-\epsilon_{\q_2}-\epsilon_{\q_3}-\hat{H}_z}\,,
\end{align}
in which $\hat{H}_z$ is the non-interacting Hamiltonian for the one-dimensional harmonic oscillator.

Evaluating the $\delta$-functions, we then obtain three coupled expressions of the form
\begin{multline} \label{eq:q2DCoupledEqs}
f_0^{2}\,\mathcal{T}_{ij}^{-1}\left[E_3-\frac32\epsilon_\q-\left(n+\frac12\right)\omega_z\right]\eta_{\q,n}^{(k)}=\\\sum_{\q',\,n'}\left[\eta_{\q',n'}^{(i)}+\eta_{\q',n'}^{(j)}\right]\sum_{l,\,l'}\frac{e^{-|\q+\q'/2|^{2}/\Lambda^{2}}e^{-|\q/2+\q'|^{2}/\Lambda^{2}}}{E_3-\epsilon_{\q}-\epsilon_{\q+\q'}-\epsilon_{\q'}-(n+l+1)\omega_z}f_{l}f_{l'}\langle n,l|n',l'\rangle\,,
\end{multline}
with the same values for $\{i,\,j,\,k\}$.  Above, the left-hand side contains the $\mathcal{T}$ matrix appearing in Eq.~\eqref{eq:q2DTMatrix} and the harmonic-oscillator wave function $f_0$ in Eq.~\eqref{eq:q2DHOEigenFunc}.  The harmonic-oscillator quantum numbers, $l$ and $n$, correspond respectively to relative atom-atom and atom-pair motion in the $z$ direction, while $\q$ is the relative atom-pair momentum in the $x$-$y$ plane (and similarly for the primed variables).

The Gaussian cutoff functions can be approximated the same way as in the 3D problem --- see Eq.~\eqref{eq:10}.  After projecting to the $s$-wave, Eq.~\eqref{eq:q2DCoupledEqs} then leads to the system of equations~\eqref{eq:q2DMatEq}-\eqref{eq:q2DMatM} in Sec.~\ref{sec:Quasi-2D_System}.

\twocolumngrid

\bibliography{trimers_paper.bib}

\onecolumngrid

\end{document}